\documentclass[aps,prc,fleqn,superscriptaddress,floatfix,twocolumn]{revtex4}

\usepackage{graphicx}
\usepackage{amsmath,amssymb,amsfonts}
\renewcommand{\imath}{i}

\begin{document}

\title{Physics With Two Time Dimensions}

\author{Jacob~G.~Foster}
\affiliation{Complexity Science Group, Department of Physics, University of Calgary, Calgary, Alberta, Canada T2N 1N4}

\author{Berndt~M\"uller}
\affiliation{Department of Physics \& Center for Theoretical and Mathematical Science, \\ 
Duke University, Durham, NC 27708, USA}

\date{\today}

\begin{abstract}
We explore the properties of physical theories in space-times with two time dimensions. We show that the common arguments used to rule such theories out do not apply if the dynamics associated with the additional time dimension is thermal or chaotic and does not permit long-lived time-like excitations. We discuss several possible realizations of such theories, including holographic representations and  the possibility that quantum dynamics emerges as a consequence of a second time dimension.
\end{abstract}

\maketitle

\section{Introduction}

Modern unified theories of the fundamental interactions generally posit that we live, at microscopic scales, in a universe with more than three spatial dimensions. Following Kaluza's \cite{Kaluza:1921} and Klein's \cite{Klein:1926} original idea, the extraneous dimensions are assumed to be compactified so that excited modes of the elementary particle fields involving nontrivial dynamics in these dimensions are exceedingly massive and therefore not directly observable. The sole exception to this rule may be gravity \cite{LargeXdim}. 

The Kaluza-Klein concept is naturally realized within superstring theory, which requires at least six additional dimensions for internal consistency \cite{String}. It has been noted in recent years that the additional spatial dimensions do not have to be aspects of a genuine space-time continuum, but could be artefacts of the algebraic structure of some underlying quantum mechanical system \cite{Matrix}. 

One common feature of most theoretical speculations of this kind is that the additional dimensions are assumed to be spatial. The reason for this restriction is that physics in the presence of more than one {\it temporal} dimension is generally considered to violate several fundamental and well established properties of nature. The two most important of these principles are causality and unitarity, but there are also concerns that additional time dimensions imply tachyonic modes and ghost fields, i.~e.\ field components with negative norm.

We will discuss the case against more than a single time dimension in Sect.~II, followed by an overview of the arguments that have led us to reconsider this case. In Sect.~III, we first review the mathematical difficulties associated with the initial-value problem for wave equations in spaces with multiple time dimensions. After explaining why a recently proposed resolution of the ill-posed nature of the initial-value problem does not work in general for interacting fields, we will argue that interacting fields without propagating time-like modes (with respect to the additional time dimension) can evade this difficulty.  In particular, we show how the difficulty can be circumnavigated if the field is thermalized in the additional time dimension. We propose a specific model of such a quantum field theory, constructed as the boundary field theory of a six-dimensional Anti-deSitter space-time with two time dimensions containing a Schwarzschild black-hole. 

In Sect.~IV we point out that, once the notion of field theories in space-times with two time dimensions is accepted as physically viable, the {\em quantum} dynamics of the reduced field theory in Minkowski space can be generated by the mechanism of micro-canonical quantization from the underlying {\em classical} dynamics of a field in a higher-dimensional space-time that contains two time dimensions. Finally, we summarize our results and present some thoughts about the possible cosmological context of a second time dimension.

\section{Multiple Time Dimensions}

\subsection{The case against multiple time dimensions}

Let us briefly review why the basic principles of causality, unitarity, and vacuum stability are thought to be violated in the presence of additional time dimensions.
\begin{enumerate}
\item
{\em Causality:} 
In space-times with at least two time dimensions, it is always possible to construct closed time-like curves \cite{CTC}. This is especially worrisome in the context of classical physics, where observers are not part of the physical system and can ``act'' in unpredictable ways. The existence of closed time-like curves implies that an observer can revisit the past and, if we accept the tenet of  ``free will'', change it in a manner that is incompatible with the already experienced future. The classic example of a resulting inconsistency is the trip back in time, where the hero unwittingly kills his own ancestor (the grandfather paradox). Obviously, anything resembling the common notion of causality cannot be maintained under such circumstances. The question of the well-posed nature of the Cauchy (initial-value) problem in space-times with closed time-like paths was investigated in detail by Aref'eva {\em et al.} \cite{Arefeva:2009bz}.
\item
{\em Unitarity:}
While violations of causality can arise both in classical and quantum dynamics, violations of unitarity are specific to the realm of quantum mechanics. Why unitarity can be violated in theories with more than one time dimension is most easily illustrated, following Yndur{\'a}in \cite{Yndurain:1990fq}, by considering the Feynman propagator of a scalar particle in three spatial and two temporal dimensions denoted by ${\bf x}$ and $(t,\theta)$, respectively. We denote the Fourier transformed variables as $({\bf k}, \omega, \chi)$.  The potential energy $V(r)$ between by a static source, located at ${\bf x}=0$ and one at $|{\bf x}|=r$ and existing at the same ``time'' $\theta$, has the form (see Appendix for details):
\begin{equation}
\label{eq:pot}
V(r) = - \dfrac{\alpha}{r} \left[ 1 + \dfrac{2\alpha}{r} \sum_{n=1}^{\infty} e^{\imath nr/L} \right]  \ ,
\end{equation}
where $L$ is the extent of the additional time dimension (considered as compact) and $\alpha$ is the effective low-energy coupling constant. 

As seen from (\ref{eq:pot}), the additional compactified time dimension renders the scalar potential generated by a static source imaginary. Yndur{\'a}in's result also applies to the Coulomb potential generated by a static charge; thus, the energy levels of the hydrogen atom would be complex, implying a violation of the conservation of probability. In other words, a compactified second time dimension leads to violations of unitarity in quantum mechanics, seemingly excluding extended space-time models of this kind. Dvali {\em et al.} \cite{Dvali:1999hn} considered a slightly different situation, where particles are localized at $\theta = \theta_0$ in the second time dimension, and only gravity propagates freely in the additional dimension.
\item
{\em Tachyons:}
Quantum field theories in space-time with more than one time dimension generally suffer from the presence of tachyonic modes. In order to see why, we again consider a scalar field in Minkowski space augmented by a fifth, temporal dimension, with the same notation for the coordinates and the conjugate momenta as above. The dispersion relation of a free relativistic particle with mass $m$ then has the form
\begin{equation}
\omega(\textbf{k},\chi) = \textbf{k}^2 - \chi^2 + m^2 \ ,
\end{equation}
which implies an imaginary Minkowski space energy $\omega$ whenever $\chi^2 > \textbf{k}^2 +m^2$. Modes in this kinematic regime grow exponentially in the physical time coordinate $t$ implying a vacuum instability. 
\item
{\em Ghosts:}
For massless vector particles, whenever the metric $g_{\mu\nu}$ contains a second temporal eigenvalue ($+1$ in our convention), the Feynman propagator $D_{\mu\nu}(k) = -\imath g_{\mu\nu}(k^2 + \imath\epsilon)^{-1}$ contains ghost modes with negative norm, which are not eliminated as dynamical degrees of freedom by the transversality condition $k_{\mu}A^{\mu}(k) = 0$. 
\end{enumerate}

\subsection{Multiple time dimensions reconsidered}

We shall now explain why these objections against additional time dimensions may not be as ironclad as generally believed. We begin by noting that the objections are based on the assumption that the higher-dimensional space-time exists in a state resembling the vacuum state, so that local excitations are discrete in energy and their classical or quantal motion is regular.  This need not be the case for an additional time dimension.  For instance, we can imagine that excitations with an (energy-like) momentum component in the direction of a second time dimension are thermally excited with a temperature $T_5$ that is much higher than any energy scale accessible in the laboratory. The dynamics in the additional time dimension would then be diffusive, not ballistic, and would thus render any deliberate attempt to revisit a targeted event in the past futile \cite{CBR}. 

In fact, for the causality argument against a second time dimension to fail, we do not have to assume true thermalization. All we need to assume is that the excitation energy with respect to the second time dimension is sufficiently high and the physical system sufficiently nonlinear, so that all trajectories involving motion in the second time dimension are dynamically unstable. Even an infinitesimal deviation from the precise initial conditions of the closed-time path would then result in exponentially large deviations of the end-point of the trajectory from its starting point. Our fortunate hero would almost certainly miss his ancestor, or more correctly, the probability for an encounter with him would be exponentially small no matter how hard he tried.

A space-time with a thermally excited second time dimension also allows us to evade the objection based on unitarity violation \cite{Foster:2003}. If we repeat Yndur{\'a}in's calculation of the propagator of a scalar particle, but in a thermal bath rather than in the vacuum, we do not encounter unitarity violations. The periodicity of the second time dimension is now along the imaginary axis, implying Matsubara frequencies $\chi_n=2\pi\imath n T_5$, and the resulting two-body potential 
\begin{equation}
V({\bf x}) = - \frac{\alpha}{r} \left[ 1 + 2 \sum_{n=1}^{\infty} e^{-nT_5r} \right]
\end{equation}
remains real as discussed in the Appendix. Corrections to the usual Minkowski space potential $- \alpha/r$ are exponentially suppressed when $T_5$ is large.  Note that for equilibrium quantities the thermal extra time dimension behaves like an extra spatial dimension with compactification length $L = 1/T_5$; thus the high temperature limit for the thermal extra time dimension is equivalent to a small compactification radius for an extra spatial dimension when only long-time averages are considered.  

Similarly, when $T_5$ is large and one is only interested in Minkowski space observables averaged over second time intervals $\Delta\theta \gg 1/T_5$, these can be calculated using the imaginary-$\theta$ formalism. For each Matsubara frequency $\chi^{(n)}$, the dispersion relation in real physical time:
\begin{eqnarray}
\omega(\textbf{k},\chi) &=& \textbf{k}^2 - \chi_n^2 + m^2 
\nonumber \\ 
&=&  \textbf{k}^2 + (2\pi n T_5)^2 + m^2
\end{eqnarray}
only yields real Minkowski space energies $\omega$. The theory thus does not contain tachyonic modes. Finally, ghost-like modes are eliminated if the modes of vector fields polarized in the $\theta$-time direction are screened.

\subsection{Other studies of multiple-time theories}

The possibility of the cosmological emergence of additional time dimensions was considered by Sakharov \cite{Sakharov:1984ir}. Aref'eva and Volovich \cite{Arefeva:1985mv} showed that the problem of ghost modes in Kaluza-Klein theories with additional time dimensions can be avoided, if the additional time dimensions reside in compact manifolds that have no Killing vector field at all. Examples of such spaces are Kummer's $K3$ manifold or, more generally, compact hyperbolic manifolds. 

Aref'eva and Volovich showed further that Kaluza-Klein type theories with additional time dimensions would generally compactify on hyperbolic manifolds. Geodesics in such spaces have stochastic behavior; as a matter of fact, geodesic motion on Riemannian manifolds with constant negative curvature serves as a tractable mathematical model for dynamical systems with strongly chaotic properties \cite{Anosov:1967}. As we argued above, this property may also ameliorate or even eliminate the problem of macroscopic causality violations.

Bars and collaborators have extensively considered theories in spaces with two time dimensions, which possess an additional local symplectic gauge symmetry connecting position and momentum variables \cite{Bars:1997xb,Bars:1998ph,Bars:2000mz}. The imposition of a gauge constraint then allows one to reduce the dynamics to a single time dimension. This avoids all undesirable aspects of theories with two time dimensions, but at the cost that the quantum field cannot, even in principle, dynamically explore the second time dimension. Nevertheless, the possible progeny from a higher dimensional space-time may help resolve certain unexplained properties of the Standard Model \cite{Bars:2006dy}.

\section{Ultra-Hyperbolic Wave Equations}

\subsection{The Cauchy problem}

The generalization of Minkowski space with $p$ spatial and $q$ temporal dimensions is denoted as ${\bf R}^{p,q}$, and we specify a vector in ${\bf R}^{p,q}$ as $(x_1,\ldots,x_p;t_1,\ldots,t_q)$. We will refer to $t_1$ as the ``physical'' time coordinate. Later we will be interested in the special case $p=3,q=2$. In this case, we denote the five-dimensional vector as $({\bf x};t,\theta)$, where $t$ represents the physical time coordinate. 

The prerequisite for a causal dynamics in any space-time manifold is that the Cauchy problem for the relevant differential equations is well posed. Let us, for simplicity, consider the scalar $n=p+q$ dimensional wave equation
\begin{equation}
\label{eq:UHDE}
\sum_{i=1}^{p} \frac{\partial^2 u}{\partial x_i^2} - \sum_{k=1}^{q} \frac{\partial^2 u}{\partial t_k^2} = 0 .
\end{equation}
Differential equations of this type with $q>1$ are called {\em ultra-hyperbolic}.  The Cauchy problem for linear ultra-hyperbolic differential equations of the type (\ref{eq:UHDE}) has been studied by various authors \cite{John:1938,Owens:1947} and explicit solutions were given by Owens \cite{Owens:1942}. The minimal conformal representations of solutions of the wave equation were constructed by Kobayashi and {\O}rsted \cite{Kobayashi:2003} for even values of $n$.  

If there is more than one temporal dimension, the initial data must be given on a $(n-1)$-dimensional hypersurface of mixed metric signature, i.~e.\ one which contains temporal directions. Without loss of generality, we assume that the initial data are given on the hypersurface $t_1=0$ and denote the vector of the remaining temporal coordinates by ${\bf t}' = (t_2,\ldots,t_q)$.  Building on a generalized mean value theorem by Asgeirsson \cite{Asgeirsson:1936}, John showed that, if the initial value problem has a solution, this solution is unique  \cite{John:1938}. 

Courant \cite{Courant:1962} pointed out that Asgeirsson's theorem also implies that the initial value problem is generally not well-posed, because a global solution of the wave equation (\ref{eq:UHDE}) does not exist for arbitrary initial conditions. This can be seen as follows \cite{Courant:1962} (see also Tegmark \cite{Tegmark:1997jg}). First, Asgeirsson's theorem implies that knowledge of the initial value data in an infinitesimal region in the temporal coordinates, 
\begin{equation}
\label{eq:disk}
t_1=0;  \qquad |{\bf t}'-{\bf t}'_0| \leq \varepsilon; \qquad |{\bf x}-{\bf x}_0| \leq a
\end{equation}
is sufficient to fix the data in the finite cone-shaped region
\begin{equation}
\label{eq:cone}
t_1=0; \qquad |{\bf t}'-{\bf t}'_0| + |{\bf x}-{\bf x}_0| \leq a 
\end{equation}
on the hypersurface.  In other words, while data on the whole {\em spatial} domain contained somewhere in the past light-cone of a space-time point is needed to uniquely determine the value of the function $u({\bf x},{\bf t})$, the data need to be known only on an infinitesimal strip in the {\em temporal} directions $t_2,\ldots,t_q$ (see Fig.~1). Because the infinitesimal disk (\ref{eq:disk}) is sufficient to determine the Cauchy data for a solution of eq.~(\ref{eq:UHDE}) over the whole cone (\ref{eq:cone}), it is not possible to prescribe arbitrary data in the cone, but only those which are compatible with the data in the infinitesimal disk. Since a solution of the ultra-hyperbolic equation is unique if it exists, this implies that no solution exists for arbitrary Cauchy data on the cone-like region. (The same argument applies to the disk-like region itself, because it is always possible to find a narrower disk which already fixes the Cauchy data on the larger disk.)

\begin{center}
\begin{figure}[h]
\label{fig:Cauchy}
\includegraphics[width=0.85\linewidth]{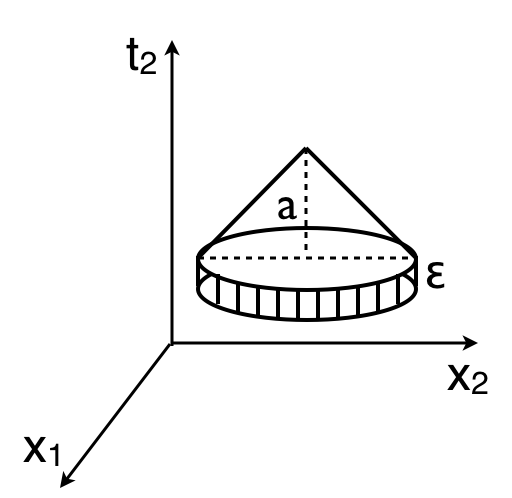}
\caption{Initial data domain for the Cauchy problem for the scalar wave equation in an space-time with two time dimensions. The spatial extent of the domain on the co-dimension one hypersurface $t_1=0$ can be chosen arbitrarily large (here as a sphere of radius $a$), while an infinitesimal extent in the second temporal dimension $t_2$ is sufficient. The data given in the infinitesimal disk completely fix the data in the entire cone-shaped region indicated in the figure. This result implies that the Cauchy problem on any finite volume on the $t_1=0$ hypersurface has generally no solution. If a solution exists, however, it is unique.}
\end{figure}
\end{center}

These results, however, leave the question unanswered which Cauchy data are admissible. Recently, Craig and Weinstein \cite{Craig:2009,Weinstein:2008aj} showed that the existence of a unique causal solution can be ensured by imposing a nonlocal constraint on the initial values. They proved that a unique global solution of the ultra-hyperbolic wave equation (\ref{eq:UHDE}) exists if the $(p+q-1)$-dimensional Fourier representation
\begin{eqnarray}
\label{eq:FT}
u_0({\bf x},{\bf t}') &=&  \int \frac{d^pk\, d^{q-1}\chi\,}{(2\pi)^{p+q-1}} \tilde{u}_0({\bf k},\chi)\,  e^{i{\bf k}\cdot{\bf x} - i \chi\cdot{\bf t}'} ,
\nonumber \\
u_1({\bf x},{\bf t}') &=&  \int \frac{d^pk\, d^{q-1}\chi\,}{(2\pi)^{p+q-1}}  \tilde{u}_1({\bf k},\chi)\,  e^{i{\bf k}\cdot{\bf x} - i \chi\cdot{\bf t}'} 
\end{eqnarray}
of the initial data 
\begin{eqnarray}
u_0({\bf x},{\bf t}') &=& u({\bf x},0,{\bf t}')
\nonumber \\
u_1({\bf x},{\bf t}') &=& \frac{\partial u}{\partial t_1}({\bf x},0,{\bf t}')
\end{eqnarray}
satisfies the constraint
\begin{equation}
\label{eq:CW}
\tilde{u}_0({\bf k},{\bf \chi}) = \tilde{u}_1({\bf k},{\bf \chi}) = 0 \quad {\rm for}~ \chi^2 > {\bf k}^2  .
\end{equation}
The physical interpretation of the constraint (\ref{eq:CW}) is that the initial conditions of the field only have support in the region of space-like momenta on the $(n-1)$-dimensional hyperplane $t_1=0$. This result is easy to understand. If we write the Fourier representation of the solution $u({\bf x},{\bf t})$ of eq.~(\ref{eq:UHDE}) as
\begin{multline}
u({\bf x},{\bf t}) =  \int \frac{d^pk\, d^{q-1}\chi\,}{(2\pi)^{p+q-1}} \left( \tilde{u}_0({\bf k},\chi) \cos[\omega({\bf k},\chi)t_1] \phantom{\frac{\tilde{u}_0}{\omega}} \right.
\\
\left.  + \frac{\tilde{u}_1({\bf k},\chi)}{\omega({\bf k},\chi)} \sin[\omega({\bf k},\chi)t_1] \right) e^{i{\bf k}\cdot{\bf x} - i \chi\cdot{\bf t}'} 
\end{multline}
with $\omega({\bf k},\chi) = \sqrt{{\bf k}^2-\chi^2}$, the constraint (\ref{eq:CW}) ensures that only real values of $\omega({\bf k},\chi)$ appear in the representation, and thus $u({\bf x},{\bf t})$ remains bounded.

The problem with the result of Craig and Weinstein is that the constraint (\ref{eq:CW}) only works for a free field, but fails to protect the global evolution of $u$ if the field is interacting, i.~e.\ if the wave equation contains nonlinear terms in $u$. For illustration, let us consider a small quadratic term in the wave equation:
\begin{equation}
\label{eq:NLUHDE}
\sum_{i=1}^{p} \frac{\partial^2 u}{\partial x_i^2} - \sum_{k=1}^{q} \frac{\partial^2 u}{\partial t_k^2} + \alpha u^2 = 0 .
\end{equation}
If $\alpha u^2 \ll u$, we can consider the nonlinear term as a perturbation, at least for a very short time interval $0 \leq t_1 \leq \tau$. The time derivative of the solution of (\ref{eq:NLUHDE}) at time $t_1=\tau$ is given by
\begin{eqnarray}
\frac{\partial u}{\partial t_1}({\bf x},\tau,{\bf t}') 
&\approx & \frac{\partial u}{\partial t_1}({\bf x},0,{\bf t}') 
+ \int_0^\tau dt_1\, \frac{\partial^2 u}{\partial t_1^2}({\bf x},0,{\bf t}')
\nonumber \\
&& \hspace{-2.2cm} 
= u_1({\bf x},{\bf t}')  
%\nonumber \\
%&& \hspace{-1cm} 
+ \int_0^\tau dt_1 \left( \alpha u_0^2 + \frac{\partial^2 u_0}{\partial{\bf x}^2}  
  - \frac{\partial^2 u_0}{\partial{\bf t}'^2}  \right) 
\nonumber \\
&&  \hspace{-2.2cm} 
= u_1({\bf x},{\bf t}') + \tau \alpha\, u_0({\bf x},{\bf t}')^2
\nonumber \\
&& \hspace{-2.1cm} 
-\, \tau \int\frac{d^pk\, d^{q-1}\chi\,}{(2\pi)^{p+q-1}}\, \omega({\bf k},\chi)^2 \tilde{u}_0({\bf k},\chi) e^{i{\bf k}\cdot{\bf x} - i \chi\cdot{\bf t}'}.
\end{eqnarray}
The Fourier transforms of the first and last term only have support in the domain $\chi^2 \leq {\bf k}^2$, if this is true for the initial data at $t_1=0$. The second term, however, has the Fourier transform
\begin{equation}
\tau \alpha \int \! \frac{d^p k'\, d^{q-1}\chi'}{(2\pi)^{p+q-1}}\, \tilde{u}_0({\bf k}-{\bf k}',\chi-\chi') \tilde{u}_0({\bf k}',\chi') ,
\end{equation}
which generally is nonzero for ${\bf k}=0$ even if $\chi\neq 0$. This consideration implies that the constraint (\ref{eq:CW}) is generally not preserved by the physical time evolution of the field in the presence of interactions.  The physical interpretation of our result is that two space-like momentum modes on the hyperplane $t_1 = 0$ can fuse into a time-like momentum mode, if their space-like momenta cancel or nearly cancel.

\subsection{Fields with quenched time-like modes}

The consideration at the end of the previous subsection shows that the initial value problem for interacting fields in space-times with more than a single time dimension can be well-posed only if time-like momentum modes are dynamically suppressed or ``quenched'', so that the time ($t_1$) evolution does not violate the Craig-Weinstein constraint (\ref{eq:CW}). In order to understand what this requirement implies, we consider the dynamics of the field within the $(n-1)$-dimensional subspace of the hyperplane $({\bf x},{\bf t}') \in {\bf R}^{p,q-1}$, on which the initial-value data are given. For specificity, let us consider the case $p=3, q=2$, i.~e.\ the extension of Minkowski space by one additional time coordinate, which we denote by $t_2\equiv\theta$. We also simply write $t_1\equiv t$. We are looking for a field theory, in which all momentum modes with $|\chi| > |{\bf k}|$ are dynamically quenched and only modes with $|\chi| \leq |{\bf k}|$ survive. If we could find such a field theory, we would not need to arbitrarily impose the constraint (\ref{eq:CW}) on the initial data, rather, the constraint would arise dynamically from the field theory. 

The thermal $N=4$ supersymmetric Yang-Mills theory in (3+1) dimensions at large number of colors $N_c$ and large 'tHooft coupling constitutes precisely such a theory. The spectral function of current and stress-energy correlation functions of this theory have been studied in detail \cite{Teaney:2006nc,Kovtun:2006pf}, and it was shown that the theory does not contain propagating modes, except the hydrodynamic mode which is only weakly damped. The hydrodynamic mode with the space-like dispersion relation $\chi^2={\bf k}^2/3$ exhibits minimal damping from a shear viscosity $\eta$ that saturates the so-called KSS bound $\eta = s/4\pi$ \cite{Policastro:2001yc,Kovtun:2004de}. 

Time-like excitations in the thermal super-Yang-Mills theory at strong coupling are rapidly damped \cite{Gubser:2008as,Chesler:2008uy}. (Note that we here refer to the $\theta$-direction as ``time'' and to $T_5 \equiv T_\theta$ as ``temperature''.) In the gauge theory, this process corresponds to the ``democratic'' splitting of the original elementary excitation into lower-energy modes, which become part of the thermal bath within a finite time \cite{Hatta:2008tx,Chesler:2008wd}. One may visualize this as ``shooting a water gun under water'', where the initially well identified stream of water emerging from the gun quickly fragments into smaller and smaller eddies and quickly loses its identity and dissipates into the surrounding water.

In order to see how the exponential growth in physical time of modes with $\chi^2 > {\bf k}^2$ can get suppressed by such rapid damping, we consider a schematic example.  Let us assume that all dynamical excitations in $\theta$-time decay like Gaussians with a time constant $\tau$, and we focus on the ``most time-like'' modes by setting ${\bf k}=0$:
\begin{equation}
u(\theta) = e^{-\theta^2/2\tau^2}  .
\end{equation}
The Fourier transform is
\begin{equation}
\tilde{u}(\chi) = \sqrt{2\pi\tau^2}\, e^{- \tau^2\chi^2/2}.
\end{equation}
The full time-dependence of a massless mode in $t$ and $\theta$ is then obtained by setting $\omega=\pm i \chi$:
\begin{eqnarray}
u(t,\theta) &=& \int_{-\infty}^{\infty} \frac{d\chi\,}{2\pi}\, \tilde{u}(\chi)\, e^{\pm\chi t -i\chi\theta}
\nonumber \\
&=& \exp\left( - \frac{(\theta \pm i t)^2}{2\tau^2} \right)
\end{eqnarray}
Averaging this expression over $\theta$ yields a constant, which is independent of the physical time $t$. In other word, the $\theta$-time dependence decouples from the physical time dependence, which will be determined by the dynamics in Minkowski space, i.~e.\ four-dimensional $({\bf x},t)$ space-time.

\subsection{Holographic theory with two time dimensions}

This picture of a strongly coupled, thermal, supersymmetric gauge theory in higher dimensions is suggestive of Witten's holographic construction \cite{Witten:1998zw} of the 4-dimensional large-$N_c$ Yang-Mills theory as the boundary field theory of a supergravity theory on AdS$_7$ space in which two spatial dimensions have been compactified on a circle. The first compactification preserves supersymmetry, while the second one is assumed to correspond to a ``thermal'' compactification, which breaks supersymmetry by requiring periodic boundary conditions for bosons and antiperiodic ones for fermions. Although this additional benefit is important if one wants to imbed the scenario of two time dimensions into string theory, it is not essential in a more general context. Witten's construction has been used to obtain estimates for the glueball spectrum and other properties of the pure Yang-Mills theory in the large 'tHooft coupling limit \cite{Csaki:1998qr,Hashimoto:1998if}. 

\begin{center}
\begin{figure}[!ht]
\label{fig:AdS6}
\includegraphics[width=0.95\linewidth]{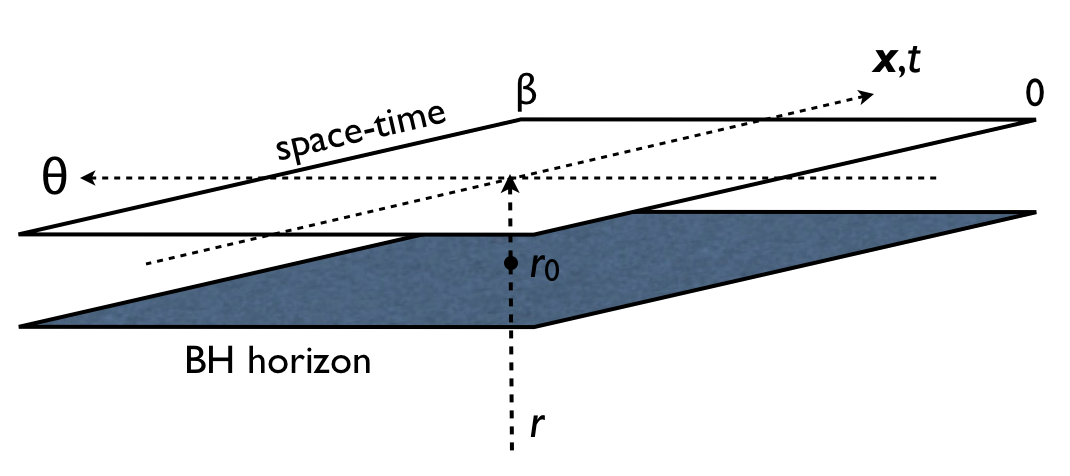}
\caption{Schematic representation of the six-dimensional Anti-deSitter-Schwarzschild geometry. The black hole is located at $r=0$; its horizon forms a black brane at $r=r_0$. The five-dimensional boundary space-time contains two time dimensions, one of which ($\theta$) is assumed to be thermalized by the black hole.}
\end{figure}
\end{center}

One possible realization of a holographic theory with two time dimensions, one of which is thermal, is six-dimensional Anti-deSitter space space AdS$_6$ containing a black hole \cite{Kuperstein:2004yf,Wen:2004qh,Cai:2007zza}. The imbedding of the black hole identifies which direction in the two-dimensional plane of time dimensions is thermal. As elsewhere in this manuscript, we label this direction by the coordinate $\theta$; the orthogonal time direction is labeled $t$. The metric of this space-time is given by
\begin{eqnarray}
\label{eq:AdS6-BH}
\ell_s^{-2}ds^2 &=& (r/R)^2 \left[ f(r) d\theta^2 - dt^2 + d{\bf x}^2 \right] 
\nonumber \\
&&+ \frac{(R/r)^2}{f(r)} dr^2 ,
\end{eqnarray}
where
\begin{equation}
f(r) = 1- \left(\frac{r_0}{r}\right)^5 
\end{equation}
is the metric factor determining the location of the black hole horizon along the radial AdS$_6$ direction. The AdS$_6$-black hole geometry is illustrated in Fig.~2. The AdS$_6$-Schwarzschild metric (\ref{eq:AdS6-BH}) is an solution of the 6-dimensional Einstein equations with a nonvanishing cosmological constant. Here $\theta$ is taken to be a periodic coordinate with period 
\begin{equation}
\beta_5 = T_5^{-1} = \frac{4\pi R^2}{5r_0} ,
\end{equation}
corresponding to the inverse Hawking temperature of the black hole. Assuming holographic duality, the AdS$_6$ metric (\ref{eq:AdS6-BH}) has been used to estimate the low-lying glueball masses, string constant, and other infrared properties of the large-$N_c$ Yang-Mills theory with qualitatively good results \cite{Kuperstein:2004yf,Wen:2004qh,Cai:2007zza}.

The holographic representation also provides an intuitive explanation for the rapid quenching of $\theta$-time modes of the strongly coupled boundary gauge theory. Such excitations are represented as string-like excitations of the dual gravity theory, which fall into the black hole encoding the thermal bath and disappear behind the horizon in a finite time \cite{Gubser:2008as,Chesler:2008uy}.

\subsection{Non-supersymmetric gauge theories}

It may not be necessary to restrict the theory to either strong coupling or supersymmetry. In fact, in an interacting thermal field theory all time-dependent excitations not protected by conserved quantum numbers ultimately decay to the equilibrium ensemble. For example, in a thermal Yang-Mills theory without matter content all time-dependent gauge field configurations are quenched and only static chromomagnetic fields survive at large times \cite{Bodeker:1998hm}. The time scale for the decay is given by the inverse temperature $T_5$ multiplied by a power (and possibly logarithms) of the coupling constant $g$ of the gauge theory.

This argument suggests that the thermal 5-dimensional Yang-Mills theory in ($3+2$) dimensions reduces to the vacuum Yang-Mills theory in ($3+1$) dimensions on length and physical time scales much larger than the inverse temperature. This statement is known to be true when the physical time dimension $t$ is analytically continued from Minkowski space ${\bf R}^{3,1}$ to Euclidean space ${\bf R}^4$. In this case, all static correlation functions in ${\bf R}^{4,1}$ are dominated at long distances by the $n=0$ Matsubara frequency $\chi_n = 2\pi i n T_5$ of the thermal unphysical time dimension $\theta$. The effective coupling constant of the 4-dimensional gauge theory is $g_4=g_5T_5$, where $g_5$ is the dimensionful coupling of the 5-dimensional gauge theory, which in non-renormalizable and thus requires a physical short-distance cut-off. In order to obtain a 4-dimensional coupling of order unity, $T_5$ cannot differ very much from this cut-off.

The physical gauge theory in Minkowski space thus emerges as the long-distance limit of a thermal 5-dimensional gauge theory in a space-time with two time dimensions. Formally, this 5-dimensional theory would be constructed by doubly analytically continuing the 5-dimensional euclidean gauge theory defined on ${\bf R}^4\times S^1$, where the compactified dimension corresponds to the unphysical time dimension, which is hidden from observation by thermal physics.

\section{Quantization by a second time dimension}

Building on the results of the Sect.~III.D, we can now argue that a {\em classical} dynamics in the $(d+2)$-dimensional space-time induces an effective {\em quantum} dynamics in the $(d+1)$-dimensional space-time. This concept is not entirely new. It is implicit in Beck's idea of chaotic quantization \cite{Beck:1995} and explicit in the chaotic quantization model of Bir{\'o} {\em et al.} \cite{Biro:2001dh}, who argued that quantized non-Abelian gauge theory can be understood as the infrared limit of the corresponding classical gauge theory in a higher space-time dimension. In some respects, our argument can also be viewed as a realization of 'tHooft's conjecture \cite{tHooft:2001fb} that quantum mechanics emerges as a limit of classical dynamics under appropriate conditions of dynamical information loss.

The idea that a classical field theory can be equivalent to a lower-dimensional quantum field theory is also inherent in the concept of gravitational holography \cite{Susskind:1994vu}, although it is usually explored only in the strong coupling limit for the most widely studied models \cite{Maldacena:1997re}. In these models the additional space-time dimension is usually taken to be spatial and endowed with a nontrivial metric. Here, we consider an additional temporal dimension, not a spatial one.

\subsection{Micro-canonical quantization}

In order to explore the quantum mechanics inducing property of an second time dimension it is useful to consider a time dimension attached to the euclidean analogue of Minkowski space-time, i.~e.\ by considering the manifold defined by the coordinates $({\bf x},t=-i\tau,\theta)$ with the metric
\begin{equation}
ds^2 = - d\theta^2 + d\tau^2 + d{\bf x}^2 .
\end{equation} 
As before, $\theta$ denotes the unphysical time coordinate.  We denote the conjugate momentum to $\theta$ (the $\theta$-``energy'') as $p_5=H$, and the momenta conjugate to $({\bf x},\tau)$ as $({\bf p},p_4)$, where $p_4$ is related to the Minkowski space energy by $E=-ip_4$.

Consider a classical dynamical system whose evolution in the unphysical time variable $\theta$ is ergodic. We denote the dynamical variables (fields) collectively by $A$ and the $\theta$-energy associated with a particular set of these variables as $H[A]$. Observed over long times $\theta$, the system will visit  all points on the hypersurface defined by $\delta(H[A]-H_0)$, where $H_0$ is the initial $\theta$-energy of our system. The long $\theta$-time average of an observable ${\cal O}[A]$ is then equal to the microcanonical average:
\begin{eqnarray}
\langle {\cal O} \rangle &=& \lim_{\theta\to\infty} \frac{1}{\theta}
  \int_0^\theta d\theta'\, {\cal O}[A(\theta')] 
  \nonumber \\
&=& V(H_0)^{-1}\, \int {\cal D}A\,\delta(H[A]-H_0)\,{\cal O}[A] ,
\label{microcan-Q}
\end{eqnarray}
where $V(H_0)$ is the volume of the constant-$H$ hypersurface. The last expression in (\ref{microcan-Q}) is the definition of the quantum mechanical average of the observable ${\cal O}$ in a formalism called {\em micro-canonical quantization} \cite{Strominger:1982xu,Iwazaki:1984kx}. Using the Fourier representation of the delta function, the micro-canonical average can be written in the form
\begin{equation}
\langle {\cal O} \rangle 
= V(H_0)^{-1}\,  \int_{-\infty}^{\infty} \frac{d\tilde\theta}{2\pi}
  \int {\cal D}A\, e^{i(H[A]-H_0)\tilde\theta}\, {\cal O}[A] .
\end{equation}

For systems with an infinite number of degrees of freedom, such as field theories, the microcanonical average is known to agree with the canonical average under very general conditions \cite{Strominger:1982xu}. When this equivalence holds, we can express the expectation value of the observable ${\cal O}$ as
\begin{equation}
\langle {\cal O} \rangle 
= Z(\beta)^{-1}\, \int {\cal D}A\,e^{-\beta H[A]}\,{\cal O}[A] ,
\label{can-Q}
\end{equation}
where $Z(\beta) = \int {\cal D}A\,e^{-\beta H[A]}$ is a partition function and the parameter $\beta$ is, as usual, defined by the condition
\begin{equation}
\beta\, \int {\cal D}A\,e^{-\beta H[A]} H[A] 
\equiv - \beta\frac{\partial Z}{\partial\beta} = \beta H_0 .
\end{equation}
Upon the identification $\beta=\hbar^{-1}$, the expression (\ref{can-Q}) becomes the usual definition of the (canonical) quantum mechanical expectation value of the observable ${\cal O}$ as a functional integral in euclidean space. $H[A]$ is recognized as the euclidean space action associated with the system. In our five-dimensional space-time $H$ is simply the energy variable associated with the additional time dimension. The quantum of action (Planck's constant) $\hbar$ appears as the quasi-temperature associated with the ergodic finite-energy trajectory of the system in the five-dimensional space-time. Note that this definition of the expectation variable does not involves a thermal ensemble, only a ($\theta$)-time average over an ergodic, classical trajectory in the five-dimensional space-time. 

\subsection{Stochastic and chaotic quantization}

The micro-canonical ensemble has been widely used in lattice gauge theory to calculate vacuum expectation values of various operators \cite{Callaway:1982eb,Namiki:1982mj,Callaway:1983ee,Iwazaki:1984sh,Iwazaki:1986cp,Fukugita:1985hq}. In practice, the micro-canonical ensemble is generated by a stochastic trajectory obeying a Langevin equation in an auxiliary ``time'' variable. This is the essence of {\em stochastic quantization} \cite{Parisi:1980ys,Namiki:1992wf,Namiki:1993fd}, which is thus recognized as a schematic implementation of micro-canonical quantization \cite{Callaway:1984pn}. 

Of course, the micro-canonical average can be generated by any other appropriate method, e.~g.\ by a deterministic, but ergodic process in an additional {\em physical} (not ``auxiliary'') time dimension. Such a process is naturally realized when the evolution of the five-dimensional system in the additional time variable $\theta$ is strongly chaotic. The method of defining the vacuum expectation value of a system by means of the ergodic average over a chaotic trajectory has been called {\em chaotic quantization} \cite{Beck:1995,Biro:2003qn}. 

An example of a physical system that thus generates its own quantum analogue in a reduced dimension is the classical lattice gauge theory \cite{Biro:2001dh}. A numerical study comparing the ``self-quantizing'' five-dimensional classical gauge theory with the canonically quantized gauge theory in four dimensions, which was undertaken by Bir\'o and M\"uller for the compact U(1) lattice gauge theory \cite{Biro:2003jq}, showed excellent agreement for the expectation value of the Polyakov loop as a function of the gauge coupling if Planck's constant $\hbar$ is identified with the product $T_5 a$, where $a$ is the lattice spacing.

\subsection{Analytic continuation}

In order to describe physics in real physical time, one must analytically continue the euclidean time coordinate $\tau$ to the real time coordinate $t=-i\tau$. For the canonical average (\ref{can-Q}), this is exactly what is usually done in quantum field theory, where the Minkowski space vacuum expectation value of an observable ${\cal O}$ is {\em defined} by the analytic continuation of the euclidean space functional integral. The critical question is whether the same analytic continuation also applies to, and makes sense for, the micro-canonical and ergodic time averages (\ref{microcan-Q}).

The analytic continuation involves a number of subtleties. First, the real-time analogue of the functional integral (\ref{can-Q}) does not involve a real integrand and is thus not mathematically well defined. For the micro-canonical average (\ref{microcan-Q}) the problem is that the analytically continued $\theta$-energy (the Minkowski space action) is not positive definite, and thus the constant-$H$ hypersurface does not have a finite volume, even for a field theory with infrared and ultraviolet cut-offs. In the stochastic average, the mathematical subtleties reside in the definition of a Langevin process for a system governed by a Minkowski space action with indefinite sign. 

There has been a limited amount of work on stochastic quantization in Minkowski space \cite{Huffel:1984mq,Nakazato:1985zj}. For certain systems, a complex Langevin process can be defined and simulated, which generates the appropriate microcanonical ensemble \cite{Callaway:1985vz,Okano:1992hp}. In other cases, the complex Langevin process can be stabilized by a suitable kernel \cite{Okamoto:1988ru}. Recently, complex Langevin evolution has been applied to generate the real-time quantum evolution of SU(2) Yang-Mills theory in (3+1) dimensions \cite{Berges:2007nr}.

\section{Summary and Outlook}

We have argued that additional time dimensions are not ruled out, if the dynamics governing the evolution of elementary fields in these dimensions is thermal or quasi-thermal. There are several distinct scenarios for this:
\begin{enumerate}
\item
If the universe is fundamentally quantum mechanical, an additional, thermalized time dimension is equivalent to a compactified spatial dimension on time scales long enough for all non-equilibrium processes to be dampened out. Because fermions and bosons obey different periodicity conditions, this scenario implies the breaking of any fundamental supersymmetry and thus could explain why supersymmetry is not realized at low energies. At distances and times much larger than $\hbar/T_5$, the presence of the additional time dimension is essentially invisible, except for the supersymmetry breaking and its influence on various low-energy constants.
\item
If the universe is fundamentally classical, an additional, highly excited and quasi-thermal time dimension will generate quantum dynamics in the four-dimensional space-time. Quasi-thermal means that the dynamics is micro-canonical globally, and the local ergodic ensemble can be approximated by a thermal ensemble. Planck's constant $\hbar$ is given by the product $T_5a$, where $a$ is the ultraviolet cut-off length of the classical field theory. In this scenario, the additional time dimension is reflected in the quantum mechanical nature of Minkowski space physics.
\end{enumerate}
Any scenario of a thermal, additional time dimension raises a number of questions.
What is the origin of the high temperature $T_5$ associated with the additional time dimension and why is $T_5$ so high? While the thermal cosmic background radiation is naturally understood as remnant of the thermalized energy density contained in the excited vacuum state that drove cosmic inflation, the temperature associated with an additional time dimension must be of a more primordial nature. It also needs to remain constant as a function of physical time; otherwise fundamental laws of nature would be observed to change over cosmic time scales.

A partial answer to this question emerges from the observation that, in the presence of two time dimensions, temperature is a two-component vector. If the observable universe is homogeneous, the physical time direction will simply be the time direction that is orthogonal to the primordial temperature vector. The ekpyrotic cosmological model \cite{Khoury:2001wf}, which posits that the Big Bang was initiated by a violent process, e.~g.\ the collision between two energetic branes, could then be re-interpreted to provide an explanation for the high temperature associated with the additional time dimension. It would be interesting to construct a model that explains why the temperature of the universe associated with the physical time dimension decreased over time, but the temperature associated with the second time dimension remained high. 

We hasten to emphasize that the physical laws will differ severely from those customarily assumed in those domains of space-time, e.~g.\ during or before the cosmic Big Bang, where the full dynamics in one of the two time dimensions is not effectively screened by thermal or quasi-thermal dynamics. For example, the law of causality will not apply in such regions, at least not the usual notions of causality. How such physical laws will break down in domains where two time dimensions are dynamically active may depend quite strongly on the details of the cosmological model.

Another question is whether and how the presence of an additional temporal, thermal dimension could be detected experimentally. With respect to $\theta$-time averages, the second temporal, but thermally excited dimension behaves just like a compactified spatial dimension with compactification length $L=1/T_5$. Experimental probes of average properties thus will require energies $E$ of order of the temperature, or precision of the order of some power of $E/T_5$. However, it may be possible to probe for the existence of real $\theta$-time dynamical phenomena, which may have a much longer duration if the dynamics of relevant fields is weakly coupled. For most field theories, the longest lived excitations are of hydrodynamic nature, because their decay is protected by local conservation laws. It may thus be possible to probe for violations of causality or unitarity induced by such dynamical processes in $\theta$-time. It would be interesting to derive a general theory of such microscopic causality violations.

{\em Acknowledgments:} This work was supported in part by DOE grant DE-FG02-05ER41367. We thank T.~S.~Bir\'o and I.~Ya.~Aref'eva for valuable comments on an earlier version of this manuscript.

\appendix*

\section{The Five-Dimensional Propagator}

Let us consider a massless scalar field with the Lagrangian
\begin{equation}
L = \dfrac{1}{2} \partial_{\mu}\phi \partial^{\mu}\phi + j\phi
\end{equation}
and its Feynman propagator
\begin{equation}
D(k) = \dfrac{\imath}{k^{2} + \imath\epsilon} \ .
\label{eq:prop}
\end{equation}
Here $k$ denotes the conjugate momenta to the spacetime coordinates, and we can break $k$ up into $\textbf{k} = (k_{x}, k_{y}, k_{z})$, $\omega$, the momentum conjugate to $t$, and $\chi$, the momentum conjugate to the second time direction $\theta$. After a Fourier transformation, the five-dimensional wave equation with a source term
\begin{equation}
\partial_{\mu}\partial^{\mu} \phi(x) = j(x)
\end{equation}
is solved by
\begin{equation}
\phi(x) = \dfrac{\imath}{(2\pi)^{5}} \int d^{5}k\, e^{-\imath k x} D(k) j(k) \ ,
\label{eq:phi-j}
\end{equation}
where $j(k)$ is the Fourier representation of the source. Following Yndur{\'a}in \cite{Yndurain:1990fq}, we consider a point source of strength $q$ at rest with respect to physical time and in the second time dimension. The source is given by
\begin{equation}
j(x) = q \int_{-\infty}^{\infty} ds\ \delta(\xi^{\mu}(s) - x^{\mu}) \ ,
\end{equation}
where $\xi^{\mu}(s)$ is the position of the point source in 5-space-time, parameterized by the proper length $s$.  If the source is fixed at the spatial origin and at $\theta=0$, we have $\xi^{\nu}(s) = (s, 0, 0 , 0, 0)$, and the Fourier transform of the source distribution is simply
\begin{equation}
j(k) = 2\pi\, q\, \delta(\omega)\ .
\end{equation}
Inserting this result into the expression (\ref{eq:phi-j}) for the potential and using the propagator (\ref{eq:prop}) yields
\begin{equation}
\phi(x) = - \dfrac{q}{(2\pi)^{4}}\int d^{3}\textbf{k}\ d\chi\ 
\dfrac{e^{\imath(\textbf{k} \cdot \textbf{x} - \chi\theta)}}
{\chi^{2} - \textbf{k}^{2} + \imath\epsilon}
\label{eq:phi-pot}
\end{equation}
We now switch to spherical coordinates and integrate out the angular parts, obtaining for the integral over $\textbf{k}$:
\begin{equation}
\int d^{3}\textbf{k}\ \dfrac{e^{\imath\textbf{k} \cdot \textbf{x}}}{\textbf{k}^{2} -\chi^{2} -\imath\epsilon}
= \dfrac{2\pi}{\imath |\textbf{x}|}\int_{-\infty}^{\infty}dk\ \dfrac{k\ e^{\imath k |\textbf{x}|}}{k^{2} -\chi^{2} -\imath\epsilon}\ .
\label{eq:contour}
\end{equation}

The integral over $k$ is easily evaluated by closing the integration contour in the positive complex half-plane, i.~e.\ only the poles with ${\rm Im}\ k > 0$ contribute. In order to perform the integral, we need to specify the values of the conjugate momentum $\chi$, which contribute. The case considered by Yndur{\'a}in \cite{Yndurain:1990fq}, which we examine first, is that the second time coordinate $\theta$ is compactified on a circle $S^{1}$ with radius $L$, i.~e.\ that $\theta$ is cyclical with period $2\pi L$. The single-valuedness of the plane wave solution in $\theta$ then implies the discrete frequencies $\chi_n = n/L$, where $n \in {\bf Z}$.  For real $\chi_n$ the contour integral (\ref{eq:contour}) becomes
\begin{equation}
\dfrac{2\pi^{2}}{|\textbf{x}|}e^{\imath |\chi_n|\, |\textbf{x}|} = \dfrac{2\pi^{2}}{|\textbf{x}|}e^{\imath |n|\, |\textbf{x}|/L}\ .
\end{equation}
Returning to the expression (\ref{eq:phi-pot}) for the field, we obtain: 
\begin{eqnarray}
\phi(x) &=& \dfrac{q}{8\pi^{2}|\textbf{x}|L}\sum_{n \in {\bf Z}} e^{\imath (|n|\, |\textbf{x}| - n \theta)/L}
\\ \nonumber 
&=& \dfrac{q}{8\pi^{2}|\textbf{x}| L} \left[ 1 + 2\sum_{n=1}^{\infty} e^{\imath n|\textbf{x}|/L} \cos \frac{n\theta}{L} \right] \ .
\end{eqnarray}
We now consider the potential energy between the static source at $\textbf{x}=0$ and another static source located at $|\textbf{x}|=r$, also at $\theta=0$. Identifying $\alpha = q^2/(8\pi^2 L)$, we reproduce Yndur\'ain's result (apart from an overall sign):
\begin{eqnarray}
V(r) &=& -q\, \phi(x) = - \dfrac{\alpha}{r} \left[ 1 + 2 \sum_{n=1}^{\infty} e^{\imath nr/L} \right] 
\nonumber \\
&=& - \dfrac{\imath\alpha}{r \tan\frac{r}{2L}} \ .
\end{eqnarray}
This shows that a compactified second time dimension introduces an undesirable imaginary component into the scalar potential generated by a static source. Because the same is true for the Coulomb potential generated by a static charge, the energy levels of, e.~g., the hydrogen atom will be complex, implying a violation of the conservation of probability. In other words, a compactified second time dimension leads to violations of unitarity in quantum mechanics, excluding extended space-time models of this kind. We note that Dvali {\em et al.} \cite{Dvali:1999hn} considered a slightly different model, where all fields except gravity are constrained to a specific point in the $\theta$-time.

Now consider the alternate case where we assume that the momentum (energy) component $\chi$ conjugate to $\theta$ is  thermalized with temperature $T_5$. The ``thermal'' average of the physical-time Feynman propagator is obtained by replacing the integral over $\chi$ with a summation over the Matsubara frequencies $\chi_{n} = 2\pi \imath n T_5$.  The evaluation of the contour integral then yields the result
\begin{eqnarray}
\phi(x) &=& \dfrac{q}{8\pi^{2}|\textbf{x}|}\ T_5 \sum_{n \in {\bf Z}} e^{(n \theta - |n|\, |\textbf{x}|)T_5} 
\\
&=& \dfrac{q\, T_5}{8\pi^{2}|\textbf{x}|} \left[ 1 
       + 2 \sum_{n=1}^{\infty} e^{-n|\textbf{x}| T_5} \cosh\, n\theta T_5 \right] 
\nonumber \\
&=& \dfrac{q\, T_5}{16\pi^{2}|\textbf{x}|} \left[ \coth \frac{|\textbf{x}|-\theta}{2}T_5 + \coth \frac{|\textbf{x}|+\theta}{2}T_5 \right] ,
\nonumber \\ \nonumber 
\end{eqnarray}
which is purely real. This shows that all corrections to the potential from the thermal dynamics in $\theta$ fall off exponentially with the distance from the source.

\end{document}